\documentclass[useAMS,usenatbib]{mn2e}
\usepackage{graphicx}
\def\sech{{\rm sech}}
\def\aap{A\&A}
\def\aaps{A\&AS}
\def\aj{AJ}
\def\apj{ApJ}
\def\apjl{ApJL}
\def\mnras{MNRAS}
\def\nat{Nature}
\def\pasp{PASP}

\title[PSF tails and galaxy halo light]{Point spread function tails and the measurements of diffuse stellar halo light around edge-on disc galaxies}
\author[Roelof S. de Jong]{Roelof S. de Jong\\%
Space Telescope Science Institute, 3700 San Martin Drive, Baltimore, MD 21218, USA}
\begin{document}

\date{Accepted 2008 May 28. Received 2008 May 15; in original form 2008 April 3}

\pagerange{\pageref{firstpage}--\pageref{lastpage}} \pubyear{2008}
\label{firstpage}

\maketitle

\begin{abstract}
  Measuring the integrated stellar halo light around galaxies is very
  challenging. The surface brightness of these haloes are expected to
  be many magnitudes below dark sky and the central brightness of the
  galaxy. Here I show that in some of the recent literature the effect
  of very extended Point Spread Function (PSF) tails on the
  measurements of halo light has been underestimated; especially in
  the case of edge-on disc galaxies.  The detection of a halo along
  the minor axis of an edge-on galaxy in the Hubble Ultra Deep Field
  can largely be explained by scattered galaxy light. Similarly,
  depending on filter and the shape one assumes for the uncertain
  extended PSF, 20 to 80 per cent of the halo light found along the minor
  axis of scaled and stacked Sloan Digital Sky Survey (SDSS) edge-on
  galaxy images can be explained by scattered galaxy light. Scattered
  light also significantly contributes to the anomalous halo colours
  of stacked SDSS images. The scattered light fraction decreases when
  looking in the quadrants away from the minor axis. The remaining
  excess light is well modelled with a S\'ersic profile halo with
  shape parameters based on star count halo detections of nearby
  galaxies. Even though the contribution from PSF scattered light does
  not fully remove the need for extended components around these
  edge-on galaxies, it will be very challenging to make accurate halo light
  shape and colour measurements from integrated light without very
  careful PSF measurements and scattered light modelling.
\end{abstract}


\section{Introduction}

In recent years we have begun to appreciate the importance of galaxy
stellar envelopes as tracers of the hierarchical galaxy formation
process. Hierarchical models in a $\Lambda$CDM context predict that
the stellar envelopes around galaxies are created from many disrupted
satellites, where size, shape, amount of substructure, and metallicity
of the envelope principally depend on the primordial power spectrum,
the reionisation epoch, the star formation history of accreted dwarfs,
and the total dark matter mass of the host galaxy
\citep[e.g.,][]{BekChi05,BulJoh05,AbaNav06,PurBul07}. The discovery of
the Sagittarius Dwarf currently being disrupted by the Milky Way
\citep{Iba94} and the highly structured envelope around M31
\citep[e.g.,][]{FerIrw02,Iba07} has given much credence to the
hierarchical model. While these observations were performed using
resolved stars, many measurements of galaxy haloes have been attempted
using integrated light
\citep[e.g.,][]{MorBor94,Fry99,WuBur02,ZibWhi04,ZibFer04}. These
observations are very difficult, as the halo light is typically at
least a factor of 10$^{4}$ below sky level, and therefore careful
attention has to be paid to flat fielding and sky subtraction. Here I
investigate another effect that has sometimes been underestimated in
integrated light studies: the effect of scattered light in extreme PSF
tails when examining edge-on galaxies.

The effect of convolving a PSF with a spherical light distribution can
be fairly well estimated. For instance, with an elliptical galaxy the
light in the central region will be dispersed by the generally broader
PSF shape. Further out, the light distribution is nearly unaffected as
the PSF shape is steeper than the slowly declining galaxy profile. By
over-plotting the PSF on the measured light distribution normalised to
the same central brightness, one gets a fairly good impression of
which radii are affected by light convolution as modelled by the PSF.

For edge-on galaxies the procedure is not as simple as the luminosity
profile perpendicular to the disc can be steeper than the outer tails
of the PSF. Close to the midplane of the galaxy the vertical light
distribution will still be modified by a point-source-like
convolution; however, farther away we are less dominated by the local
light. Instead the light scattered from the whole disc --- not just
the central core--- significantly contributes to the measured
distribution at large scale heights. To estimate this contribution at
more than 2 scale lengths above the disc (about 6--15 scale heights)
one should not use a PSF profile normalised to the central surface
brightness of the edge-on galaxy, but instead normalised to the total
brightness of the galaxy. Calculating the actual scattered light
contribution at any point is obviously best determined by convolving a
2D model of the intrinsic light distribution with a full 2D PSF.


Another problem arises when studying galaxy haloes from stacked galaxy
images, scaled to a common size, in order to reach fainter levels. One
can estimate the PSF of the stack by combining appropriate stellar
images from each field, scaled by the same factor as the galaxy in the
frame. However, for a typical sample selection the distribution of
scaling is not symmetric and highly biased toward the scales near the
selection limit of the sample. The distribution of scaled PSF images
is therefore strongly skewed, meaning that combining the PSFs using
median or mean values gives very different results. Assuming that the
galaxies are very similar after scaling, no such skewed distribution
is present and median and mean combining should give very similar
results.

\begin{figure*}
\includegraphics[width=\linewidth,trim=1.5cm 0.5cm 1.5cm 0.5cm,clip=true]{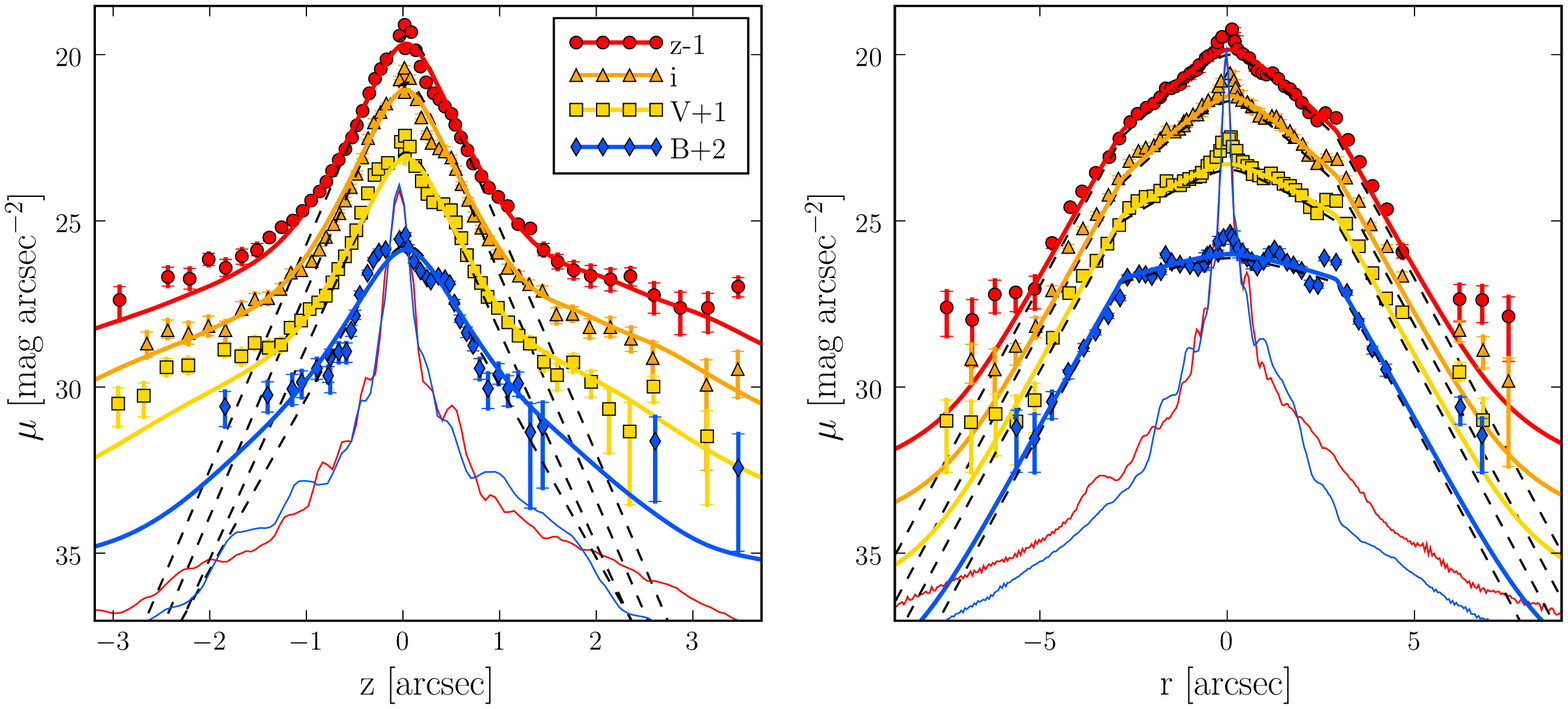}
\caption{Minor (left) and major (right) axis surface brightness
  profiles of the edge-on HUDF galaxy. The data points are derived
  from \citet{ZibFer04}. The profiles for the different bands are
  coded in different colours and symbols and are offset in order to
  avoid confusion, as indicated by the legend. The input galaxy model
  is shown by black dashed lines for the different passbands. Thin
  solid lines indicate crosscuts through the $B$ and $z$-band PSFs
  (arbitrarily normalised), whilst thick lines show the convolved
  galaxy model using the same colours for the different passbands as
  used for the data points.
\label{UDFprofs}
}
\end{figure*}

In this research note I investigate a few cases in the literature
where the effects of PSF convolution with edge-on galaxies has been
underestimated. I will also show that the colour measurements of the
envelopes around galaxies can be strongly affected by scattered
light. In particular, in Section~\ref{HUDF} I will show the effect of
PSF convolution on the light distribution around an edge-on galaxy in
the Hubble Ultra Deep field \citep[HUDF;][]{BecSti06} as measured by
\citep{ZibFer04}. In Section~\ref{SDSS} I estimate the scattered light
effect on the stacked SDSS image analysis of
\citet{ZibWhi04}. Finally, in Section~\ref{Concl} I summarise my
conclusions and give a few recommendations.


\section{Hubble Ultra Deep Field galaxy}
\label{HUDF}
Comparing the PSF and minor-axis galaxy profile normalised to the same
central brightness leads to an underestimate of the contribution from
PSF scattered light. However, this method has appeared in the
literature and been used to argue that scattered light will not
contribute to the halo light measurement \citep[e.g.,][]{Fry99,
ZibFer04}. \citet{Fry99} were unable to detect a halo around NGC\,4244
to their limiting surface brightness and therefore their discussion of
scattered light was rather moot. (The NGC\,4244 halo has now been
detected by resolved star count measurements;
\citealp{Seth07,deJ07}).  \citet{ZibFer04} do however detect an extra
component around a disc galaxy in the HUDF and report anomalous
colours for this extended component.

The HUDF galaxy studied by \citet{ZibFer04} is a highly inclined disc
galaxy at a redshift of about 0.32 with at best a small
bulge. \citet{ZibFer04} carefully removed the contribution from
contaminating (mainly background) sources around the galaxy and
extracted the luminosity profiles in wedges along the major and minor
axes. They determined profiles for all HUDF HST/ACS passbands (F435W,
F606W, F775W, F850LP, termed $B$, $V$, $i$, and $z$ hereafter for
simplicity) and these data points are reproduced in
Fig.\,\ref{UDFprofs}.

To estimate the contribution from scattered light at large radii I use
a standard disc galaxy model \citep[e.g. as used by][]{vdK88}. I
represent the radial distribution with an exponential disc projected
edge-on, which accurately matches the inner profile. However, the
observed distribution shows a clear break in exponential scale length
at 2.9 arcsec independent of wavelength. Such breaks have been seen
many times before \citep[e.g.,][]{vdK79,PohTru06}. The light beyond
the break I model with another exponential distribution, but with
shorter scale length. The inner scale length varies from $\sim$2.3
arcsec in $B$ to 0.9 arcsec in $i$ and $z$. The scale lengths beyond
the break do not vary with wavelength within the uncertainties. I
model the vertical distribution with a $\sech$ light distribution,
which is a reasonable compromise between the theoretical $\sech^2$
distribution and the more centrally concentrated exponential-like
distribution observed in the near-infrared \citep{deGPel97}. Choosing
a different vertical profile does not significantly change the
results. Major and minor axis profiles of this model before PSF
convolution are shown as dashed lines in Fig.\,\ref{UDFprofs}.

To calculate the observed distribution from the model distribution the
PSF needs to be determined out to a radius of about 10 arcsec. Such an
extended PSF cannot be accurately measured from the HUDF itself as
there are too few bright stars. I therefore used the TinyTim HST PSF
modelling software to create artificial PSFs for each passband. These
model PSFs were checked against some bright stars in a number of unrelated F606W
and F814W images and found to match the outer profiles accurately to
at least 5 arcsec. The model galaxy was rotated with respect to the
model PSFs by approximately the same amount as the real
galaxy. Cross-cuts through the centre of the TinyTim PSFs in the
direction of the galaxy major and minor axes are shown for the F435W
and F850LP filters in Fig.\,\ref{UDFprofs}.
Comparison between the two filter profiles shows that the PSFs are
quite similar within 0.8 arcsec, but at larger radii they begin to
diverge. This disparity can lead to artificial colour gradients when
the convolved light distribution is significantly contaminated or
dominated by scattered light.

Finally, I show in Fig.\,\ref{UDFprofs} the model galaxy convolved
with the PSFs for all passband as thick solid lines. The extended
minor-axis profile seen at heights greater than 1 arcsec from the
midplane in the $i$ and $z$-bands can be fully explained by scattered
light from the inner disc. On the southeast (negative $z$) side there
is a small excess in the $V$-band that might be real, for instance due
to small blue background galaxies seen here, but this component would
have very remarkable colour properties (e.g., $V$--$i$$<$0). It could
also be due to the asymmetry in the main disc light distribution not
modelled here or to some imperfections in the PSF models at that
scale. The $B$-band shows excess light beyond my model on both sides
of the disc that could be a real halo detection, but these data points
are very close to the sky background limit and would again result in
some remarkable colours, this time mainly on the northwest
side. Clearly, a large fraction of the light seen on the minor axis in
excess of the $\sech$ distribution is due to scattered light from the
extended PSF.

The major axis profiles tell a different story. There seems to be a
clear excess of light beyond 6 arcsec radius compared to the truncated
disc model. The PSF models would need to be significantly wrong to
explain this excess. The points beyond 6 arcsec radius have large
uncertainties, but assuming they are correct, we cannot tell from
these measurements whether this light is due to a more spherical halo,
or due to the disc changing scale length yet again. It does however
show that, in cases where scattered light may be significant, the best
place to look for a (flattened) halo is, maybe counter-intuitively,
not along the minor axis, but in the four quadrants away from the
major and minor axes. These are the areas where the contribution from
the main disc and from scattered light will be smallest.

\citet{ZibFer04} also presented colour profiles of this HUDF galaxy
and argued that the halo had anomalous colours starting at 1 arcsec
above the disc, i.e., the radius where the scattered light
contribution starts to become significant. While the observed colour
profiles could in principle be checked against the convolved model, it
turns out that the expected colour profiles depend critically on the
details of the calculated PSFs (especially the diffraction spikes that
are hard to model) and the assumed galaxy light and colour
distribution. The galaxy has clear small scale colour variations due
to dust extinction and local star formation that is not incorporated
into the model. Therefore, the colour gradients observed in the main
disc are not predicted, even though the PSF can induce some colour
gradients at scale height less than 1 arcsec.  Additionally, the
observed $i$--$z$ colour gradient cannot be explained in detail here,
as the PSFs only deviate significantly beyond 2 arcsec. Unfortunately,
both the galaxy and PSF models lack enough detail to make firm colour
predictions on such small scales, especially because I do not have the
background galaxy mask to select exactly the same halo areas. Because
the light at large radii is so much dominated by scattered light, even
small changes in PSFs, galaxy model, or masks can lead to large
changes in derived halo colour. The models are, however, consistent
with the data, given the (large) errors of the observations and taking
these limitations into account. Scattered light most likely contributes to
the observed anomalous colours.

\section{SDSS image stacking}
\label{SDSS}


\citet{ZibWhi04} combined 1047 images of edge-on galaxies to reach
$\mu$$\sim$31 $r$-mag arcsec$^{-2}$, with similar depths in the $g$ and
$i$-band. The SDSS images were scaled, using a characteristic scale size
for the galaxies, and combined using an approximation for the mode of
the pixel values in the image stack. 
The resulting galaxy light distribution was modelled by a double
exponential disc (exponential in radial and vertical direction)
convolved with the effective PSF, derived from similarly stacked star
images. An extra component in addition to the convolved galaxy model
was clearly detected that followed a power-law radial light
distribution with an axial ratio of about 0.6. However,
\citet{ZibWhi04} may have underestimated the effective PSF at large
radii, probably due to the procedure used to stack the stellar images
and due to the way they extrapolated the PSF profile to large radii.


\citet{ZibWhi04} selected their edge-on galaxy sample from the
``Large-Scale Structure Sample 10'' of galaxies compiled by
\citet{BlaSch05} from the SDSS in April 2002 (about the size of SDSS
DR2). Their sample selection was based on cuts by minimum luminosity
and isophotal diameter as well as an isophotal flattening
criterion. Visual inspection removed further unsuitable candidates,
reducing the first cut sample of 1221 galaxies to a final sample of
1047. After removal of contaminating sources the galaxy frames were
rotated to align the major axis and scaled radially to a common characteristic
size in the $i$-band. The final halo results were nearly independent
of whether Petrosian, half-light, isophotal radii or exponential scale
lengths were used as reference, and I will use Petrosian radii from
here on. As a common reference for radial scaling
\citeauthor{ZibWhi04} used the median scale size of the sample as
expressed in pixels, where each pixel is 0.396 arcsec in the SDSS. The
images were combined using an approximation of the mode of all pixel
values in the stack at a given position:
\begin{equation}
{\rm approximate\ mode} = 3\times{\rm median} - 2\times{\rm average,}
\label{modeappr}
\end{equation}
where the median and average images were created from the median and
(sigma clipped) average values of the stack values.

\citet{ZibWhi04} followed the same procedure to calculate an effective
PSF for the image stacks. They present their final stacked PSFs for 4
passbands as modelled by a Gaussian core and an exponential
tail. However, their PSF models are presented out to only 15--22
pixels (depending on filter), while their observations stretch to
about 100 pixels. The raw PSF profiles are not shown and we cannot be
sure whether the exponential tail is a good approximation at radii
larger than 20 pixels. I have therefore recreated the stacked PSFs in
order to calculate galaxy models out to the last measured point.

\subsection{Determining SDSS PSFs}

The largest galaxies in the sample are several times larger than the
median value used for scaling reference and so both the galaxy and associated
PSF will be shrunk by a large factor before stacking. Thus, to create
a stacked PSF out to 100 pixels, as used in the galaxy stacks, we need
to determine the PSF on the individual frames to even larger radii. To
accomplish this, we need to combine the profiles of both saturated and
unsaturated stars. From the sample used by \citet{ZibWhi04} I selected
at random 20 galaxies and downloaded their full SDSS frames in all
passbands. From these frames I extracted the azimuthal median
radial luminosity profiles of a bright but unsaturated star and the
brightest isolated star, which was always saturated at the core. By
matching these profiles in the high signal-to-noise overlap region I
reproduced one very extended PSF profile, often reaching 200 pixels
in radius and 16 magnitudes below peak brightness.

\begin{figure}
\includegraphics[width=\linewidth,trim=.5cm 0.5cm .5cm 0.5cm,clip=true]{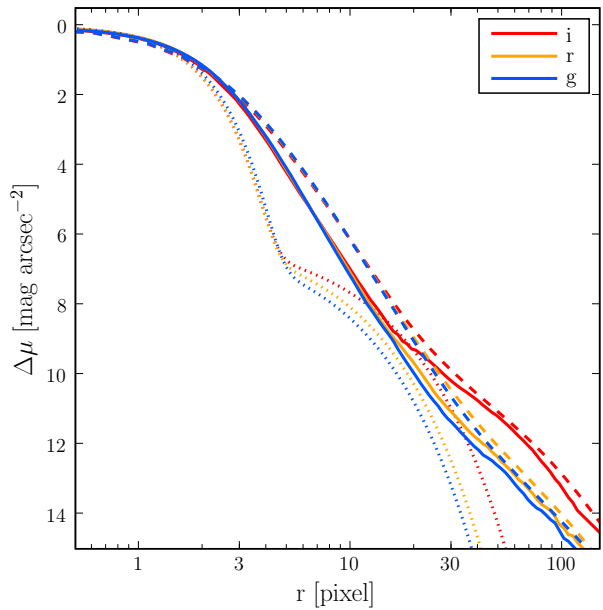}
\caption{SDSS Point Spread Functions colour coded for the $g$, $r$, and
  $i$-bands as indicated by the legend. Solid lines show typical SDSS
  PSFs derived from faint and bright stars as described in the
  text. Dashes lines show the average PSFs after scaling and
  stacking. The dotted lines show the stacked PSFs as presented by
  \citet{ZibWhi04}.
\label{PSFs}
}
\end{figure}

Comparing all 20 stellar profiles by normalising the flux within 10
pixel radius, I find they are well described by a central,
Gaussian-like core surrounded by a faint envelope
(Fig.\,\ref{PSFs}). The width of the core depends on camera column and
time, i.e.\ the camera optics and the blurring by atmospheric
seeing. The surrounding envelope is very similar in shape across all
frames, with a brightness that depends only on the total brightness of
the star. This halo is most likely caused by scattering off of dust
particles on the telescope optics. The shape and brightness of the
halo is the same in all filters with a near power law profile of slope
about -2.6. The only exception is the $i$-band filter, which shows an
additional exponential component dominating between 35 and 180
pixels. This feature is typical for backside illuminated thinned CCDs
as described in more detail in for instance \citet{SirJee05}. This
extra PSF component in the $i$-band filter explains the red haloes seen
around all bright stars in SDSS colour images, irrespective of the
star's intrinsic colour. The additional $i$-band scattered light is,
as I will show below, important when measuring colours of diffuse
haloes around edge-on galaxies.

I determine the final radial PSF profile for each filter by taking the
weighted mean of all star profiles (irrespective of central core
variations). The weights were set by the square root of the relative
brightness of the bright stars used to determine the outer halo of the
PSF. This results in PSF profiles with S/N$>$2 out to about 180
pixels. This radial extent is still not enough, as the scale sizes of
some galaxies are 6 times larger than the median value and for these
galaxies the PSF will be scaled down to approximately 30 pixels
radius.
I chose to extend the profiles beyond 180 pixels with a power law of
slope -2.6 as seen between 40 and 180 pixel radius. Choosing for
instance an exponential profile with a sharper cut off does not make a
significant difference to the final stacked PSF.

\subsection{PSF stacking and effective PSFs}
%

As far as possible I have reproduced the procedure of \citet{ZibWhi04}
to create effective PSFs. I applied the same selection criteria to the
Large-Scale Structure Sample 10 (minus eye inspection),
used the Petrosian radii to scale the 1D PSFs radially, and used the
central galaxy luminosities to scale the PSFs in intensity. The radial
scaling was done so as to conserve surface brightness, not total
luminosity of the stellar PSF profile, as was done for the galaxies.

\begin{figure*}
\includegraphics[width=0.33\linewidth,trim=.9cm 0.9cm 0.9cm 0.9cm,clip=true]{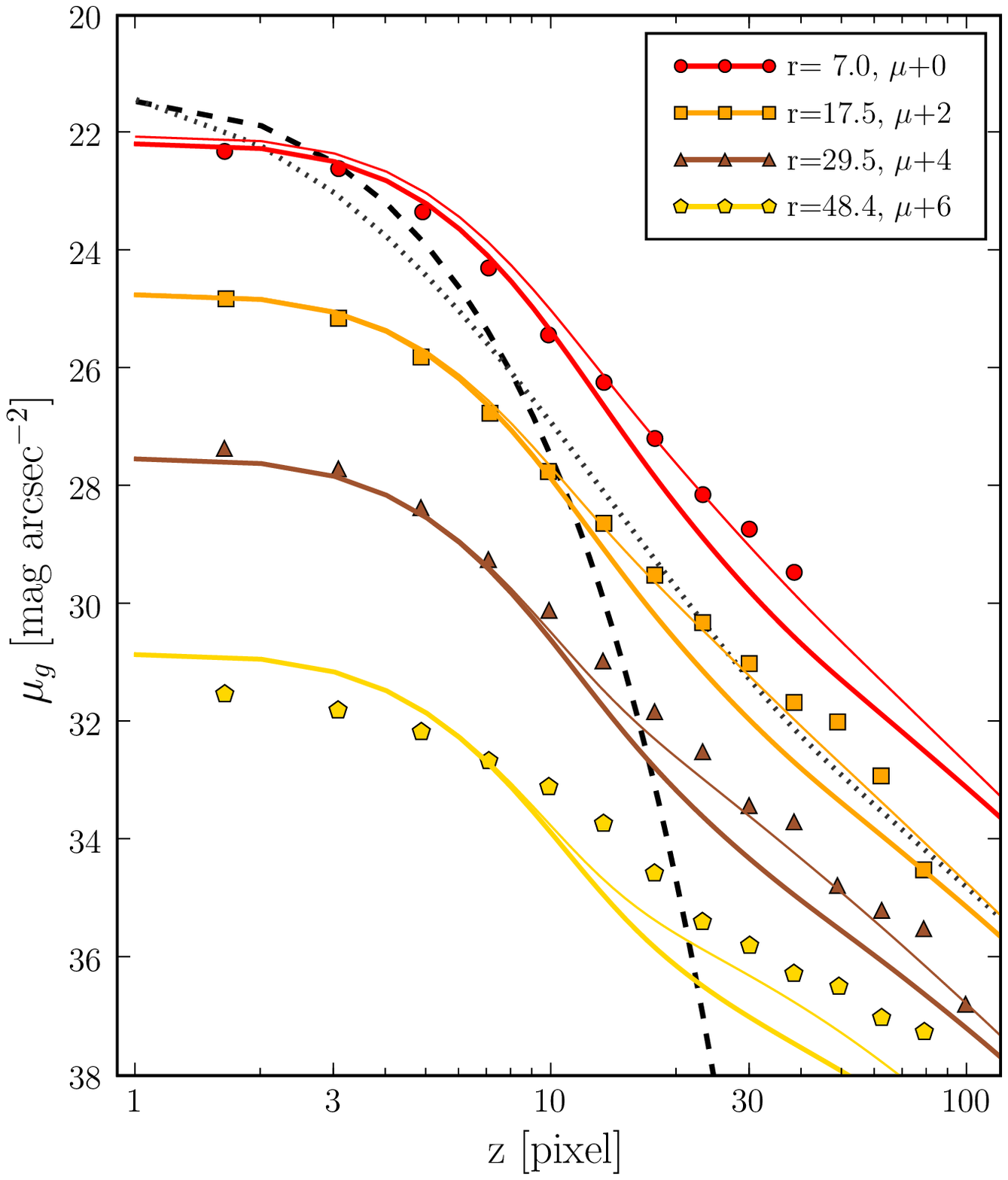}
\includegraphics[width=0.33\linewidth,trim=.9cm 0.9cm 0.9cm 0.9cm,clip=true]{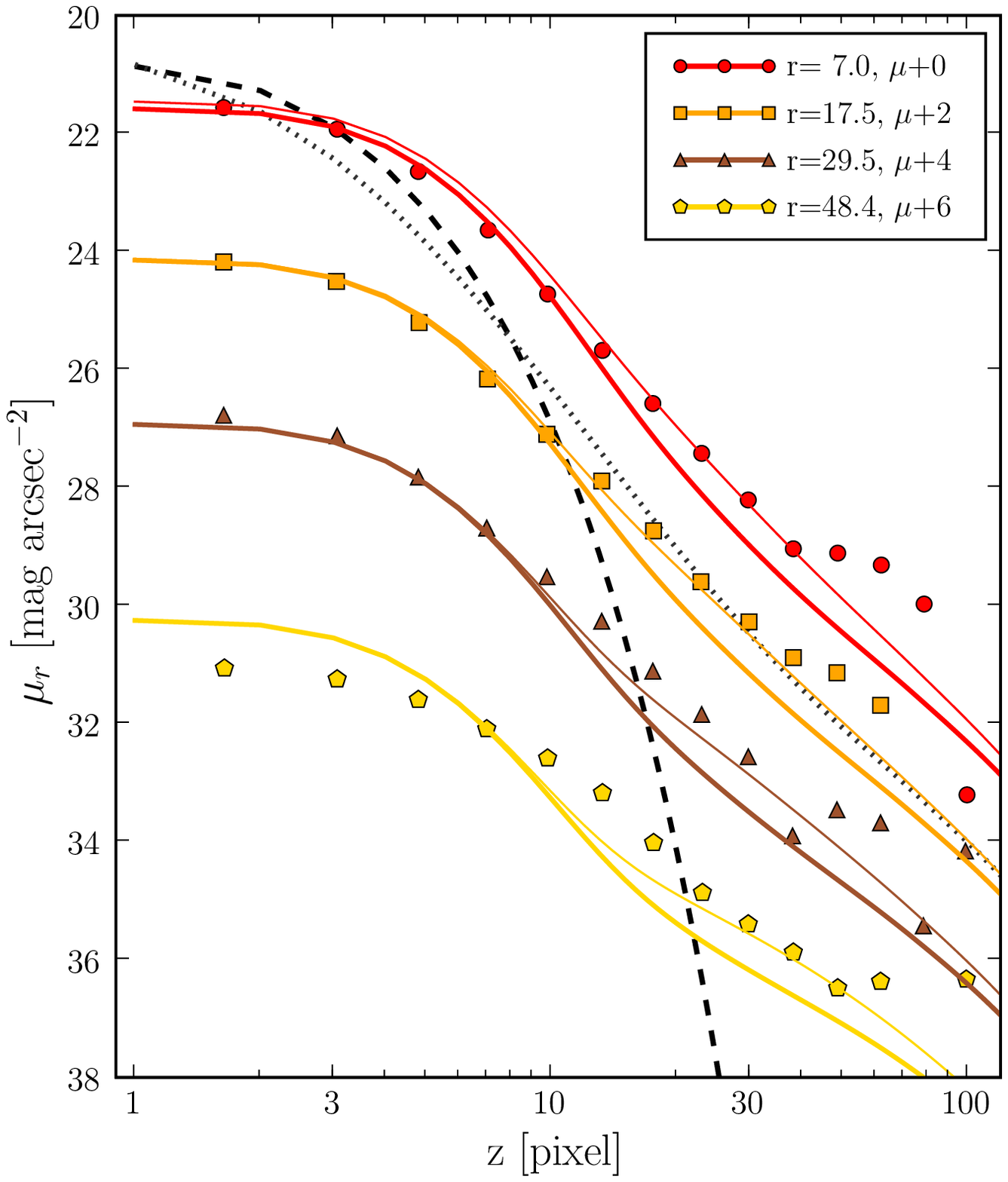}
\includegraphics[width=0.33\linewidth,trim=.9cm 0.9cm 0.9cm 0.9cm,clip=true]{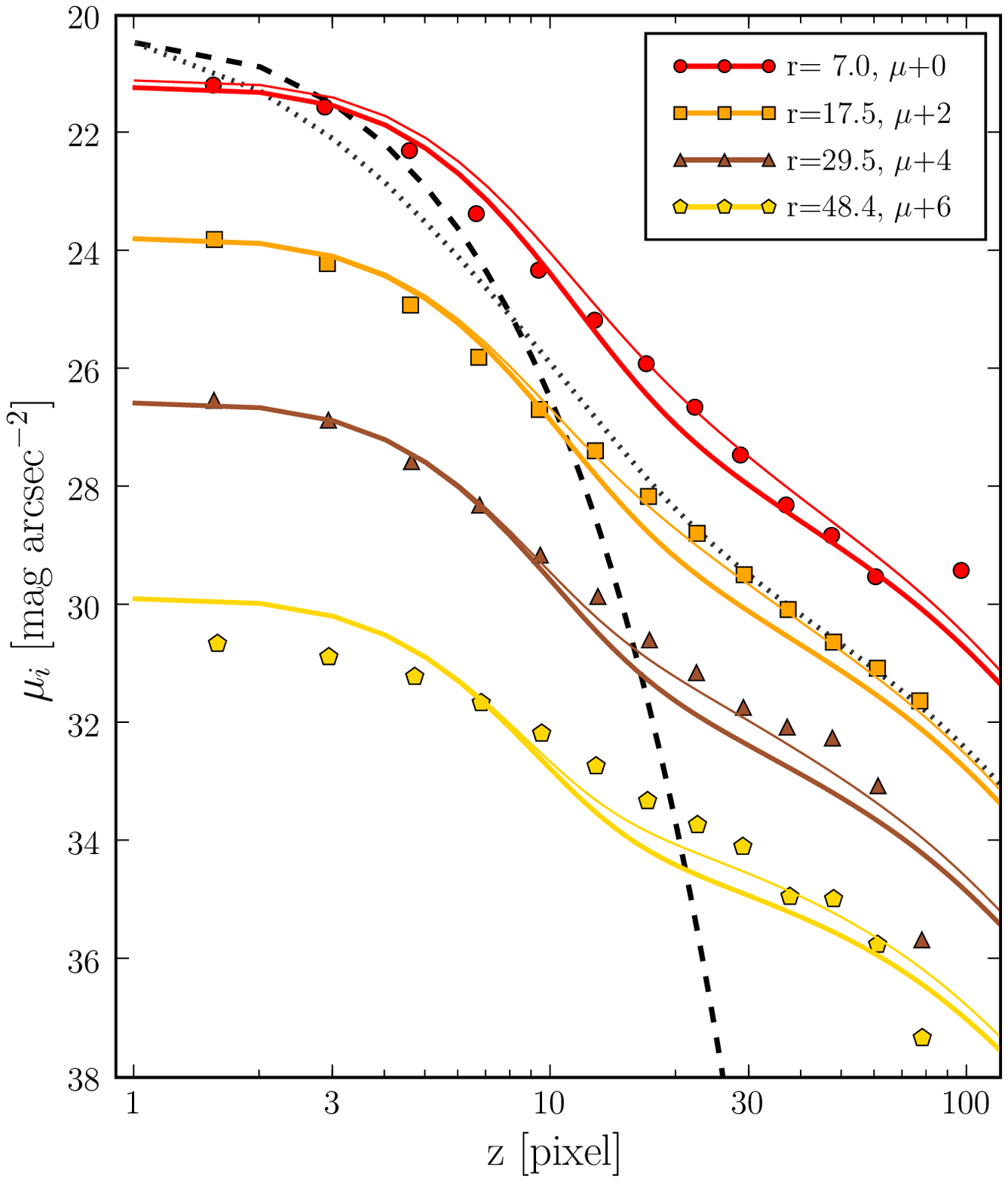}
\caption{Light profiles parallel to the minor axis at four distances
  from the galaxy centre for the $g$-band (left), $r$-band (middle),
  and $i$-band (right). Symbols represent the data as presented by
  \citet{ZibWhi04} with distances from the centre and offsets in
  surface brightness indicated by the legends. The dashed line shows
  the minor axis light profile of the galaxy model before convolution.
  The dotted line shows the average stacked SDSS PSF. The thick
  coloured solid lines show the vertical cuts at the four different
  radii through the convolved disk only model, the thin lines have an
  added S\'ersic profile halo as described in the text. 
\label{mincuts}
}
\end{figure*}

When I combine these scaled PSFs with the mode approximation of
Eq.\,\ref{modeappr}, I get negative values at most radii. The
empirically derived mode approximation of Eq.\,\ref{modeappr} only
produces reasonable results if the value distribution is not too
skewed \citep{KenStu77}. While this is probably true for the scaled
galaxy stack (assuming that the galaxies are self-similar), this is
not true for the scaled stars in the absence of noise. The apparent
scale size distribution of the galaxies is strongly skewed to small
angular sizes, and hence the scaled stellar profile intensities will
be strongly skewed, exacerbated by the steep profile shape. PSFs
stacked with mode approximation are clearly a poor representation of
the true effective PSF.
Using the median intensity at each radius instead results in a profile
nearly identical to the input PSF. This is to be expected as
everything is scaled to the median characteristic galaxy scale size
and just as many PSFs are scaled smaller as are scaled larger. The
intensity scaling is a small effect compared to the radial scaling due
to the steepness of the PSF profile. Using the average value at each
radius results in a profile that is a smoothed version of the original
PSF, with some of the central light distributed to larger radii.

Figure\,\ref{PSFs} shows typical SDSS PSFs for the $g$, $r$, and
$i$-bands (which are, as mentioned before, virtually identical to the
median stacked PSFs), together with the average stacked PSFs and the
stacked PSFs as determined by \citet{ZibWhi04} using the mode
approximation. The average stacked PSFs are clearly more extended than
the original (and hence median stacked) PSFs. However, the
\citet{ZibWhi04} stacked PSFs are much more compact than the typical
PSFs of SDSS images, especially between 3 and 10 pixels and at radii
larger than 20 pixels. This is unlikely to happen when stacking
radially scaled PSFs with conservation of surface brightness. 
The compactness of their PSFs at small radii therefore seems to be the
result of the mode approximation that is incorrect for such skewed
intensity distributions. The sharper cutoffs of their PSFs at larger
radii is due to the exponential extrapolation where a power law seems
more appropriate.  It is not entirely clear which effective stacked
PSF method is most representative of the final galaxy stack. The only
correct way of doing this is to create models of each individual
galaxy, convolved with the image PSF, and stack these models. This is
beyond the scope of this paper and I will use the averaged stacked
PSFs as
the best representation of the final galaxy stack PSF.
In the analysis in the next section I will indicate how galaxy halo
measurements are affected by using a different choice of PSF (i.e., median
stacking or \citet{ZibWhi04} stacking).

\subsection{Modelling SDSS stacked galaxy profiles}

In Fig.\,\ref{mincuts} I show the data from figure~6 of
\citet{ZibWhi04}, which shows the luminosity profiles of the stacked
galaxy images parallel to the minor axis at four distances from the
centre. I model these light distributions with a similar model galaxy
as used for the HUDF case, with an exponential radial distribution and
a $\sech$ vertical distribution. In this case I do not include a break
in the radial luminosity distribution, even though I will show it is
necessary to explain the light profiles. For simplicity I use the same
scale length and height for all filters, even though the fit could be
somewhat improved by varying the scale lengths with wavelength. The
minor axis input light profile is shown, as well as the effective PSFs
used to convolve the models. Finally, I show profiles of the convolved
2D model at each of the four vertical cuts.

The models show that significant amounts of light are found at large
radii due to PSF scattering. This is especially true for the $i$-band,
where a large fraction of the light at heights greater than 15 pixels
above the midplane could be due to scattered light. The cuts farther
away from the centre and profiles in the $g$ and $r$-passbands have
more light in excess of the scattered light profiles, but are still
often within a magnitude (a factor $\sim$2) of the model
profile. Between 15 and 40 pixels above the midplane, where the data
are most reliable, about 50\% of the light comes from scattered in $g$
and $r$, increasing to about 80\% in $i$. Clearly, any accurate
modelling of this excess light requires a thorough understanding of
the scattered light profile. The most successful detection is most
likely to occur in the four quadrants away from the major and minor
axes. The model over-predicts the luminosity near the midplane of the
profile farthest from the centre, whilst matching the data near the
midplane for the other three radii. A break in the radial profile is
necessary between the third and the fourth profiles to explain the
light distribution. Such a break will somewhat increase the
significance of the halo detection in the outer most profile.

Using the median effective PSFs, the model profiles are still within 1
mag of the data. The median stacked PSF yields a slightly larger dip
between 10 and 30 pixels, which is smoothed out when using average
stacking. The \citet{ZibWhi04} model PSFs do not explain the light at
radii larger than 30 pixels as they cut off exponentially, but at
these larger radii their effective PSF is unlikely to be correct.

The remaining difference between the observations and the convolved
disk model can be explained by adding a halo of a form that is typical
of what is found by the GHOSTS survey \citep[][de Jong et al., in
preparation]{deJRad07}. The GHOSTS survey measures halo shapes from
HST star counts not effected by PSF convolution. The typical halo
envelope can be parameterised by a \citet{Ser68} profile:
 \begin{equation}
\Sigma_{\rm B}(r) = \Sigma_{\rm eff} {\rm e}^{-b_n([r/r_{\rm eff}]^{1/n}-1)},
\label{expbul}
 \end{equation}
with $n$=5.5, the effective radius enclosing half the luminosity
($r_{\rm eff}$) equal to a third of the disk scale length, and the
effective surface brightness at this radius ($\Sigma_{\rm eff}$) equal
to a tenth of the central surface brightness of the disk. The halo has
a flattening of vertical-over-major axis of 0.65. The result of adding
this additional halo component to the model is shown by the thin solid
lines in Fig.\,\ref{mincuts}. This shows that the remaining excess can
indeed be explained by a halo profile that is typical for nearby
massive disk galaxies. However, due to the uncertainty in the PSF
convolution it will be very challenging to invert this process and
derive halo parameters from the integrated light measurements.

\begin{figure}
\includegraphics[width=\linewidth,trim=.5cm 0.3cm 0.5cm 0.3cm,clip=true]{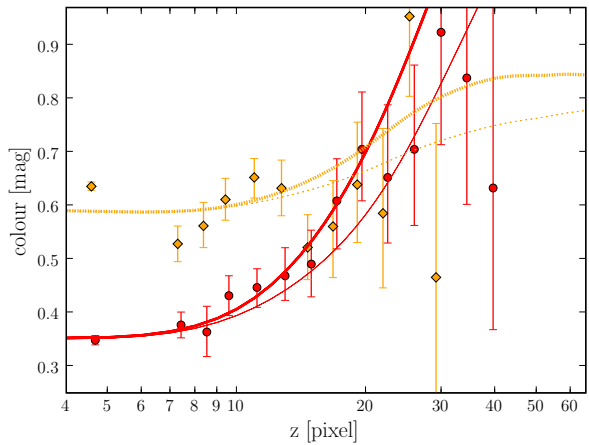}
\caption{The minor axis colour profiles. The symbols with errorbars
  show the $g$--$r$ (orange diamonds) and $r$--$i$ (red circles) data
  as presented by \citet{ZibWhi04}. The PSF convolved disk only models are
  represented by thick orange dotted ($g$--$r$) and red solid ($r$--$i$)
  lines, thin lines have an added S\'ersic profile halo as described
  in the text. 
\label{colmin}
}
\end{figure}

\citet{ZibWhi04} found that the detected halo light showed very
anomalous colours, especially in $r$--$i$. In Fig.\,\ref{colmin} I
reproduce their minor axis colour profiles with my PSF convolved
model. The colour gradient seen in $r$--$i$ can be fully ascribed to
the difference in PSF between $r$ and $i$, as the intrinsic light
distribution has no colour gradient at all. The anomalous extra
scattered light component in the $i$-band PSF causes the strong colour
gradient. The very good agreement between $r$--$i$ model and data
strongly suggests that the average stacked PSF is a fair approximation
for the effective PSF. The model also predicts a colour gradient in
$g$--$r$ between 10 and 35 pixels. No clear trend is seen in the data,
but the errorbars at these radii are so large that the model is
consistent with the data. On the other hand, only about 50\% of the
light is due to scattering in these passbands according the model,
therefore the observed colour gradient is expected to be less than the
model prediction if there is no intrinsic colour in the
galaxy. Indeed, adding the typical GHOSTS S\'ersic halo to the
models as described above reduces the colour gradients, relatively
more in $g$--$r$ than in $r$--$i$. This shows that the scattered light
still dominates the colours in $r$--$i$ for a reasonable halo profile.
%

Using the median stacked PSFs instead
of the average stacked PSFs yields model colour gradients on the minor
axis that are much steeper beyond 10 pixels than observed. Such a
steep gradient would need to be compensated by a significant real halo
component with no colour gradient. The \citet{ZibWhi04} PSFs produce a
very steep colour gradient starting at 5 pixels and a real halo would
need to dominate at these radii to reduce the scattered light colour
effect.

\section{Conclusions}
\label{Concl}

Measuring the exact halo light distribution around edge-on disc light
from the integrated light distribution is very challenging. In
addition to careful flat fielding and removal of fore- and background
sources, particular attention has to be paid to the effect of
scattered light from the central disc. Specifically, for edge-on
galaxies it is not sufficient to compare the galaxy minor axis light
profile with the PSF profile; at large heights above the disc
scattered light is contributed from the whole galaxy, not just the
centre, and a full convolution of a 2D model is needed to assess the
scattered light contamination. Basic modelling suggests that the minor
axis may not be the optimum place to detect a halo, but for a
flattened halo one should look in the quadrants away from the major
and minor axis where scattered light contribution from the disc is
smaller relative to the halo.

I find that the extended component seen along the minor axis of an
HUDF edge-on galaxy can almost fully be explained by scattered
light. The observations of \citet{ZibFer04} for this galaxy do show an
extended component along the major axis that cannot be explained by
scattered light. Similarly, the halo detection around stacked SDSS
edge-on galaxies reported by \citet{ZibWhi04} has a significant
scattered light contamination, especially along the minor axis and in
the $i$-band. Depending on the assumed best method to create a stacked
PSF the contamination along the minor axis in the $i$-band can amount
to 80\% of the light, in the $g$ and $r$-bands 50\%. The contamination
decreases with other PSF assumptions and going to one of the
quadrants, but the PSF uncertainties make it hard to derive accurate
stellar halo properties. The excess light seen on top of the disk only
model is consistent with a S\'ersic law halo with parameters typical
of those found by the GHOSTS survey \citep[][de Jong et al., in
preparation]{deJRad07}.

The anomalous minor axis halo colours reported by \citet{ZibFer04} and
\citet{ZibWhi04} are consistent with originating from scattered light
contamination. In particular, the colour gradients reported by
\citet{ZibWhi04} can be modelled by an extended PSF as
calculated from the average of the observed scaled SDSS PSFs. There
seems to be no need to invoke extreme stellar population to explain
these colours as proposed by \citet{ZacBer06}. The expected
contribution from a typical GHOSTS halo is so small in the $i$-band
that it has only a small effect on the predicted colour gradient.

There have been other reports of anomalous galaxy stellar halo
colours. \citet{LeqFor96,LeqCom98} report thick disk or halo {\em
  BVI}-band colours of edge-on galaxy NGC\,5907 that are only
consistent with colours of elliptical galaxies, i.e.\ an old, metal
rich stellar population. In this case the errors are unlikely to be
due to scattered light; the thick disk is detected at about a factor
10$^3$ below the central brightness at about 80 arcsec above the
midplane, while the PSF is already 10$^7$ below peak brightness at 16
arcsec radius. Not even a factor 10 enhancement of the scattered light
effect due to the edge-on disc configuration can make up that
difference. Similarly, the red haloes reported around Blue Compact
Galaxies \citep[see e.g., ][and references therein]{ZacBer06} are not
caused by scattered light. The galaxies are too large compared to the
PSF and the haloes show substructure different from the central galaxy
that is unlikely to be due to the substructure in the PSF.

Unfortunately, the examples of \citet{ZibFer04} and \citet{ZibWhi04}
are close to the limit where scattered light is no longer significant
and slightly larger objects would have avoided problems. The HUDF does
not contain a larger edge-on galaxy that would render scattered light
unimportant. However, the SDSS study could now be repeated on a sample
of larger galaxies as the area surveyed by SDSS has nearly tripled
since the \citet{ZibWhi04} study. Ideally, such a study would use only
a small range of scale sizes to avoid large spatial scaling
corrections and uncertainties in the effective PSF. Taking about a two
times as large an isophotal selection limit would probably suffice,
especially in the $g$ and $r$-bands.

While measuring halo properties from integrated light remains plagued
with large uncertainties, we do have a technique that allows us to
accurately measure the stellar envelopes around galaxies. By
performing star counts of resolved stellar populations we can measure
equivalent surface brightnesses to very faint limits as spectacularly
demonstrated for M31 and M33 \citep[e.g.,][]{Iba07}. These
observations are not affected by scattered light and only minimally by
flat fielding errors. The main limitations of this method lie in fore-
and background contaminating objects and low number statistics (too
few brights stars at very low surface brightnesses/densities). Very
few massive, highly inclined galaxies can be studied with ground-based
resolution, but HST allows a much larger sample of galaxies to be
studied. Indeed, preliminary results from the GHOSTS survey show
conclusively that most large galaxies do have very extended stellar
envelopes, with envelope size likely depending on galaxy mass and
bulge-to-disc ratio \citep[][de Jong et al., in preparation]{deJRad07}.

\section*{Acknowledgements}

I thank Stefano Zibetti, Annette Ferguson, and Simon White for very
constructive comments on an earlier version of this manuscript that
improved its final quality. I am grateful to Michael Blanton for
providing a machine readable version of the SDSS Large-Scale Structure
Sample 10. I thank David Radburn-Smith for a critical reading of a
previous version of this manuscript. Eric Zackrisson provided feedback
that improved my appreciation of the intricacies of the red halo
phenomenon. I acknowledge Eric Bell for making this paper see the
light of day.

Funding for the SDSS and SDSS-II has been provided by the Alfred
P. Sloan Foundation, the Participating Institutions, the National
Science Foundation, the U.S. Department of Energy, the National
Aeronautics and Space Administration, the Japanese Monbukagakusho, the
Max Planck Society, and the Higher Education Funding Council for
England. The SDSS Web Site is http://www.sdss.org/.

The SDSS is managed by the Astrophysical Research Consortium for the
Participating Institutions. The Participating Institutions are the
American Museum of Natural History, Astrophysical Institute Potsdam,
University of Basel, University of Cambridge, Case Western Reserve
University, University of Chicago, Drexel University, Fermilab, the
Institute for Advanced Study, the Japan Participation Group, Johns
Hopkins University, the Joint Institute for Nuclear Astrophysics, the
Kavli Institute for Particle Astrophysics and Cosmology, the Korean
Scientist Group, the Chinese Academy of Sciences (LAMOST), Los Alamos
National Laboratory, the Max-Planck-Institute for Astronomy (MPIA),
the Max-Planck-Institute for Astrophysics (MPA), New Mexico State
University, Ohio State University, University of Pittsburgh,
University of Portsmouth, Princeton University, the United States
Naval Observatory, and the University of Washington.


\label{lastpage}

\end{document}